\documentclass[aps,prd,10pt,twocolumn,superscriptaddress,nofootinbib,floatfix]{revtex4-2}

\usepackage{amsmath}
\usepackage{amssymb}
\usepackage{amsfonts}
\usepackage{graphicx}
\usepackage{booktabs}

\usepackage{xcolor}

\IfFileExists{newtxtext.sty}{
  \usepackage[T1]{fontenc}
  \usepackage{newtxtext}
  \usepackage{newtxmath}
}{
  \usepackage{mathptmx}
}

\usepackage{amsmath,amssymb,bm}
\usepackage{graphicx}
\usepackage{booktabs}
\PassOptionsToPackage{hyphens}{url}
\usepackage[colorlinks=true,linkcolor=blue,citecolor=magenta,urlcolor=blue]{hyperref}
\AtBeginDocument{\let\hbar\hslash}

\usepackage{tikz}
\usetikzlibrary{shapes.geometric,shapes.symbols}
\DeclareRobustCommand{\orcidicon}{
  \begin{tikzpicture}
    \definecolor{orcidgreen}{HTML}{166B3A}
    \draw[orcidgreen, fill=orcidgreen] (0,0)
    circle [radius=0.16]
    node[white] {{\fontfamily{qag}\selectfont \tiny ID}};
  \end{tikzpicture}
  \hspace{-2mm}
}

\foreach \x in {A, ..., Z}{
  \expandafter\xdef\csname orcid\x\endcsname{\noexpand\href{https://orcid.org/\csname orcidauthor\x\endcsname}{\noexpand\orcidicon}}
}

\begin{document}

\title{Dark-Matter--Gravitational-Wave Coupling in a Perfect Fluid Dark Matter Black Hole: Lensing Estimate and Ringdown Detectability Analyses}

\author{Yashmitha Kumaran\,\orcidB{}}
\affiliation{Instituto de Astronom\'ia, Universidad Nacional Aut\'onoma de M\'exico, \\ AP 70-264, Ciudad de M\'exico 04510, M\'exico.}

\date{\today}

\begin{abstract}
Two independent observational channels are examined for detecting a genuine coupling between a dark matter halo and a passing gravitational wave, for a black hole embedded in perfect fluid dark matter: an imprint on weak gravitational lensing, and an imprint on the black hole's own ringdown spectrum. For the lensing channel, a first estimate is obtained by substituting the perfect-fluid-dark-matter effective mass into the existing vacuum polar-wave deflection formula, a mass-dressing approximation that bounds the dark-matter--wave cross-term by the wave strain itself, placing it beyond the reach of current or near-future interferometric facilities. The general master equation governing a polar wave perturbation on an arbitrary static, spherically symmetric background is then derived, validated against the known vacuum solution, and specialised to the perfect-fluid case. A modest, percent-level correction to the perturbation amplitude is obtained for impact parameters up to roughly fifty gravitational radii, consistent in sign and order of magnitude with the mass-dressing estimate once an earlier, ill-posed far-field boundary condition is replaced by a finite anchor outside the horizon; the horizon itself is identified as a genuine singular point of the governing equation, so the fully regular solution, and with it the final observable deflection angle, is left for future work. For the ringdown channel, the same perfect-fluid environment is shown to produce an axial quasinormal-mode shift of two to thirty-eight per cent across the halo strengths considered, confirmed independently in the time domain, and resolvable by LISA for supermassive remnants with final mass between one hundred thousand and ten million solar masses. Black hole spectroscopy is thereby identified as the more clearly established of the two detection routes, while the lensing channel is placed on a rigorous footing without its final observable signature yet being settled.
\end{abstract}

\maketitle

\section{INTRODUCTION} \label{sec:intro}

Black holes embedded in dark matter halos and subject to a passing gravitational-wave background offer two routes to probing the microphysical state of dark matter through its coupling to gravitational radiation: an imprint on the weak-field bending of light, and an imprint on the black hole's own oscillation, or ringdown, spectrum. Several strands of the literature bear on the individual pieces of this problem, but, until now, not on the coupling itself (\S\ref{ssec:litgap}).

For a black hole surrounded by perfect fluid dark matter -- the Kiselev-type density profile that yields the exact metric function $f(r) = 1 - 2M/r + (\lambda/r)\ln(r/\lambda)$ -- the weak deflection angle has been derived via the Gibbons--Werner method through the Gauss--Bonnet theorem by \cite{Xu:2018} and \cite{Haroon:2019}, and carried to second order by \cite{Gao:2023ltr,Qiao:2023}. Separately, the weak deflection of light by a Schwarzschild black hole perturbed by a continuous gravitational wave -- using the exact Xanthopoulos \cite{Xanthopoulos:1981} polar perturbation of the vacuum Einstein equations -- has been treated with the same formalism by \cite{Pantig:2024}.

What has not been examined is what happens when both effects act on the same system at once: whether a Perfect-Fluid-Dark-Matter (PFDM) halo alters how a gravitational wave imprints itself on light bending, and, if so, whether the resulting cross-term is an observationally meaningful signature or a parametrically negligible correction. A companion question, motivated by the same physical setup, is whether the same dark matter environment leaves a distinguishable and, importantly, detectable imprint on the ringdown spectrum of the black hole itself, through the axial gravitational perturbation sector.

The lensing question is first approached with a simple, physically motivated shortcut: substituting the PFDM effective mass into the established vacuum wave-deflection formula of \cite{Pantig:2024} (\S\ref{ssec:massdressing}), evaluating the locally felt mass at the deflection zone rather than re-deriving the polar perturbation equations on the true dark-matter background. This gives a cross-term parametrically bounded by the wave strain amplitude itself, several orders of magnitude below the reach of any current or near-future Very-Long-Baseline Interferometry (VLBI) facility. Since such a substitution is an approximation rather than a derivation from the field equations, it is treated as a plausibility estimate, motivating a derivation of the general linearised field equations governing a polar wave perturbation on an arbitrary static, spherically symmetric background -- not restricted to vacuum -- validated against the known Xanthopoulos vacuum solution, and specialised to the perfect-fluid background as in \S\S\ref{ssec:master}--\ref{ssec:pfdm-coupling}. Once an initial, ill-posed far-field boundary condition is replaced by a finite anchor comfortably outside the horizon (\S\ref{ssec:pfdm-coupling}), this yields a modest, percent-level correction to the metric perturbation amplitude -- consistent, in both sign and order of magnitude, with the mass-dressing estimate, at $\lambda=0.01M$, below the $1\sigma$ Event Horizon Telescope limit for Sgr~A$^*$ \citep{Vagnozzi:2023}. A genuine singular point of the governing equation at the horizon remains to be resolved before the corresponding deflection angle can be stated as a fully regular, observable quantity; this open problem, and its consequences for the VLBI non-detectability conclusion suggested by the first estimate, are discussed in \S\ref{ssec:reconcile}.

For the ringdown channel in \S\ref{sec:qnm}, a comparatively complete result is obtained: the same perfect-fluid environment produces a shift in the axial quasinormal-mode (QNM) spectrum ranging from about $2\%$ to $38\%$ across the literature-grounded range of halo strengths considered here, and, using the LISA sensitivity curve of \cite{Robson:2019} calibrated against the ringdown signal-to-noise benchmark of \cite{Berti:2006}, this shift is shown to be resolvable for supermassive black hole remnants in the range $10^5$--$10^7\,M_\odot$.

The PFDM strength parameter $\lambda$ is not treated as a free dial in this work: three values are adopted, spanning the range of Event Horizon Telescope shadow limits on Sgr~A$^*$ \citep{Vagnozzi:2023}, from a value below the $1\sigma$ limit to a value slightly above the $2\sigma$ limit, with the most commonly adopted value in the perfect-fluid-dark-matter lensing and quasinormal-mode literature in between \citep{Qiao:2023,jana:2026}. Shadow studies of other perfect-fluid-dark-matter spacetimes adopt different ranges \citep{Hou:2018avu,Huang:2026xgt}.

The remainder of the paper is organised as follows. \S\ref{sec:formalism} summarises the established building blocks, i.e., the PFDM background, the Gibbons-Werner method, the Xanthopoulos vacuum wave perturbation, and the universal spin-$s$ perturbation potential used for the ringdown channel. \S\ref{sec:lensing} presents the lensing channel in the order: the mass-dressing estimate, its limitations, the general master equation and its specialisation to perfect fluid dark matter, and a reconciliation of the two. \S\ref{sec:qnm} derives the axial quasinormal-mode shift on the perfect-fluid background, verifies it against the standard Schwarzschild benchmark, confirms it in the time domain, and presents the LISA detectability forecast. \S\ref{sec:conc} discusses the two results together and their implications for using black holes as dark matter probes.

\subsection{Literature gap} \label{ssec:litgap}
The Gauss--Bonnet reformulation of gravitational lensing introduced by \cite{Gibbons:2008} has since been applied to black holes embedded in a wide range of exotic and dark-matter-like environments, including extended-uncertainty-principle black holes \cite{Kumaran:2020}, magnetically charged non-singular black holes \cite{Kumaran:2023}, and a broader class of black hole and wormhole spacetimes \cite{Kumaran:2021rgj}. Independently, the same formalism has been used to derive the deflection induced by a passing gravitational wave on the vacuum Schwarzschild background \cite{Pantig:2024}, and, on the dark-matter side, to obtain the PFDM deflection angle at leading and second order \cite{Haroon:2019,Gao:2023ltr,Qiao:2023}. The ringdown sector has followed a parallel but separate track: quasinormal-mode shifts have been proposed as an indirect dark matter detection channel for scalar and electromagnetic perturbations of a PFDM black hole \cite{Jusufi:2020}, and axial gravitational perturbations have since been examined for a variety of related dark matter and dark energy density profiles \cite{liang:2025qnm,Liang:2025,jana:2026,becar:2026}. What is absent from this body of work is any treatment of the two channels together, on the same background, sourced by the same halo parameter: no existing lensing study includes a gravitational wave, and no existing quasinormal-mode study addresses the corresponding lensing signature. This gap is the subject of the present work.

\section{FORMALISM} \label{sec:formalism}
This section summarises the mathematical setup.

\textbf{\textit{Perfect-fluid-dark-matter background: }}
The PFDM black hole is a static, spherically symmetric solution of the Einstein equations sourced by an anisotropic fluid with $\rho = -p_r = \lambda/(8\pi r^3)$, $p_\theta = p_\phi = -\lambda/(16\pi r^3)$ \cite{Kiselev:2003,Xu:2018}. The metric is
\begin{equation}
\mathrm{d}s^2 = -f(r)\,\mathrm{d}t^2 + \frac{\mathrm{d}r^2}{f(r)} + r^2\,\mathrm{d}\Omega^2, \quad
f(r) = 1 - \frac{2M}{r} + \frac{\lambda}{r}\ln\!\left(\frac{r}{\lambda}\right),
\label{eq:fPFDM}
\end{equation}
adopted throughout without re-derivation. It is convenient to write $f(r) = 1 - 2m(r)/r$ with $m(r) = M - (\lambda/2)\ln(r/\lambda)$, a form used both in \S\ref{sec:lensing} (mass-dressing and the master-equation specialisation) and \S\ref{sec:qnm} (the spin-$s$ potential).

\subsection{Gibbons--Werner / Gauss--Bonnet weak deflection}
For a static, spherically symmetric optical metric, the weak deflection angle of a photon with impact parameter $b$ is obtained from the Gaussian curvature $\mathcal{K}$ of the optical geometry
\begin{equation}
\hat\alpha = -\iint_{D_\infty} \mathcal{K}\,\mathrm{d}S,
\label{eq:GBT}
\end{equation}
integrated over the region exterior to the unperturbed light ray \cite{Gibbons:2008}. Applied to \eqref{eq:fPFDM} alone --- to second order in $M/b$ and first order in $\lambda/b$ i.e. in the thin-dark-matter limit --- this gives the standard PFDM deflection angle
\begin{equation}
\hat\alpha_{\rm PFDM}(b) = \frac{4M}{b} + \frac{15\pi M^2}{4b^2} - \frac{\lambda}{b}\left[1 + 2\ln\!\left(\frac{b}{2\lambda}\right)\right]
\label{eq:alpha-pfdm}
\end{equation}
of \cite{Gao:2023ltr}, which is the thin-dark-matter limit of \cite{Qiao:2023} and reduces at leading order to \cite{Haroon:2019}; it is adopted here.

\subsection{The vacuum polar wave perturbation}
\cite{Xanthopoulos:1981} obtained an exact solution of the linearised vacuum Einstein equations describing a polar perturbation of the Schwarzschild metric, oscillating at frequency $\sigma$ and expanded in Legendre polynomials $P_\ell(\cos\theta)$:
\begin{align}
A(t,r,\theta) &= f(r)\big[1+\varepsilon\, x(r) P_\ell(\cos\theta)\cos(\sigma t)\big], \\
B(t,r,\theta) &= f(r)^{-1}\big[1+\varepsilon\, y(r) P_\ell(\cos\theta)\cos(\sigma t)\big], \\
C(t,r,\theta) &= r^2\big[1+\varepsilon\big(z(r)P_\ell + w(r)P_\ell''\big)\cos(\sigma t)\big],\\
D(t,r,\theta) &= r^2\sin^2\theta\big[1+\varepsilon\big(z(r)P_\ell+w(r)\cot\theta\, P_\ell'\big)\cos(\sigma t)\big]
\end{align}
with (for vacuum Schwarzschild $f=1-2M/r$):
\begin{align}
x_0 = pq,\quad y_0=3Mq,&\quad z_0=q(r-3M),\quad w_0=rq, \nonumber\\
q =\frac{\sqrt f}{r^2},&\quad p=M-\frac{M^2+\sigma^2 r^4}{r-2M}.
\label{eq:xanthopoulos-solution}
\end{align}
\cite{Pantig:2024} used this solution in the Gibbons--Werner construction, evaluated at the equatorial plane ($\theta=\pi/2$, $\ell=2$), to obtain a closed-form finite-distance weak deflection angle,
\begin{widetext}
\begin{equation}
\hat\alpha_{\rm GW}(b,t) =  - \frac{MP(2\sigma^2b^2-3)\cos(\sigma t)\,\varepsilon}{b^3}\left(\frac{1}{u_R}+\frac{1}{u_S}\right) \\ 
- \cos(\sigma t)\,P\,b\,\sigma^2\,\varepsilon\!\left[\operatorname{artanh}\sqrt{1-b^2u_R^2} \right. \\
\left. +\operatorname{artanh}\sqrt{1-b^2u_S^2}\right] + \frac{4M}{b},
\label{eq:alphaGW}
\end{equation}
\end{widetext}
where $u_{R,S}=1/r_{R,S}$ are the inverse observer/source distances and $P$ is an $O(1)$ Legendre-polynomial-derived prefactor. This master formula was verified numerically and found to be well behaved and convergent across five orders of magnitude in $b/M$; it is taken here as a verified input with its underlying Regge--Wheeler--Zerilli-type construction.

\subsection{Universal spin-\textit{s} perturbation potential (ringdown)}
For the ringdown channel of \S\ref{sec:qnm}, the generalised Regge--Wheeler-type potential for a spin-$s$ field on any static, spherically symmetric background of the form $f(r)=1-2m(r)/r$ is used,
\begin{multline}
V_{s\ell}(r) = \\
f(r)\left[\frac{\ell(\ell+1)}{r^2} + \frac{2m(r)(1-s^2)}{r^3} - \frac{2m'(r)(1-|s|)}{r^2}\right]
\label{eq:Vsl}
\end{multline}
which reduces to the standard Regge--Wheeler potential for $s=2$ (axial gravitational) and to the Klein--Gordon potential for $s=0$ when $m(r)\to M$ \cite{Konoplya:2011}. Substituting the PFDM $m(r)$ from \eqref{eq:fPFDM} into \eqref{eq:Vsl} with $s=2$ gives the axial potential used in \S\ref{sec:qnm}; this is confirmed to reduce to the tabulated Schwarzschild benchmark values \cite{Berti:2009} in the $\lambda\to0$ limit before being used for the perfect-fluid-shifted spectrum.

\section{LENSING ESTIMATE}
\label{sec:lensing}

\begin{figure*}[t]
\centering
\includegraphics[width=0.99\textwidth]{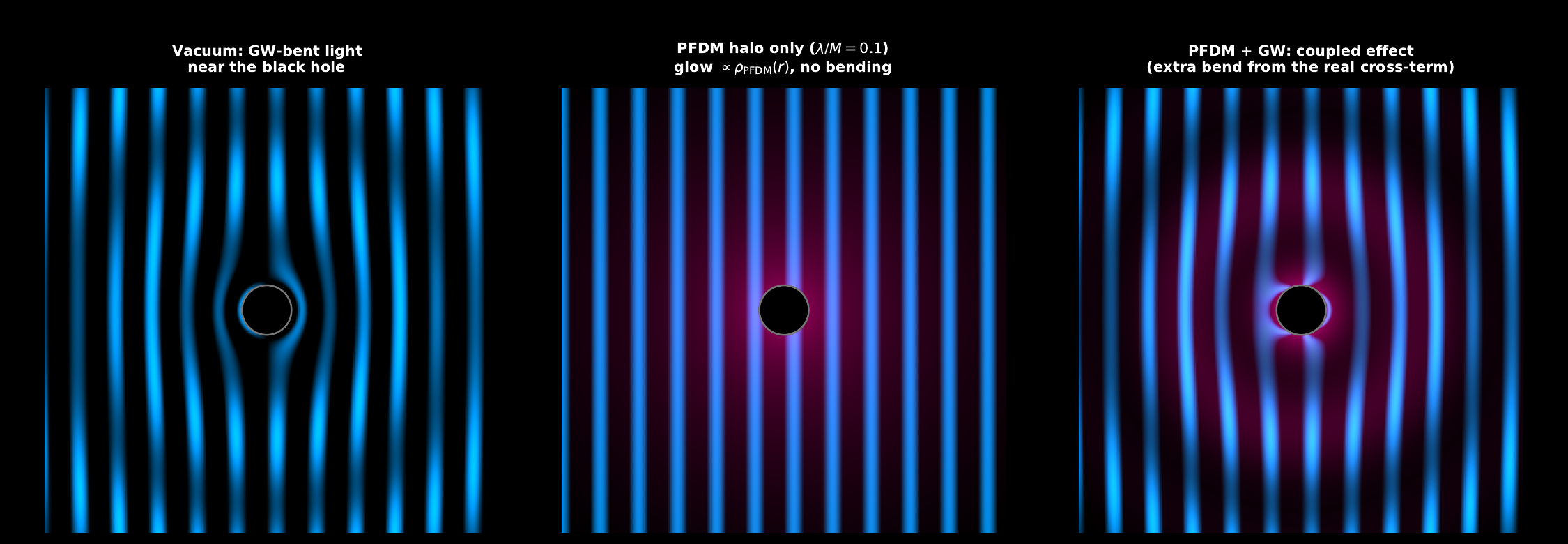}
\caption{Schematic visualisation of the three regimes considered in this section, not to observational scale: a vacuum Schwarzschild black hole under a passing polar wave; a PFDM black hole with no wave; and the combined case. The panels are purely illustrative: wavefront displacement in each panel is set from the corresponding $\hat\alpha(b)$ formula of \S\ref{sec:formalism}--\S\ref{sec:lensing}, but rescaled for visual clarity instead of being plotted to a common physical scale, and the third panel's extra warp uses an illustrative amplitude for the cross-term --- its radial shape is the real mass-dressing formula of \S\ref{ssec:massdressing}. The purple glow indicates the PFDM halo density profile. This figure is intended as a conceptual orientation aid, not a data plot.}
\label{fig:schematic}
\end{figure*}

\subsection{Choice of values of the PFDM strength parameter}
Throughout this section and \S\ref{sec:qnm}, three values of $\lambda/M$ are used (Table~\ref{tab:lambda}): a small value below the $1\sigma$ upper limit from Event Horizon Telescope shadow imaging of Sgr~A$^*$ \citep{Vagnozzi:2023}, the value most commonly adopted as an illustrative choice across the perfect-fluid-dark-matter lensing and quasinormal-mode literature, and a value slightly above the $2\sigma$ upper limit from the same analysis, taken as a generous upper edge. Other shadow studies of Sgr~A$^*$ and M87$^*$ adopt different ranges, depending on the spacetime and spin assumed \citep{Hou:2018avu,Huang:2026xgt}.

\begin{table*}[t!]
\centering
\caption{Choice of $\lambda/M$ values grounded in literature used throughout this work.}
\label{tab:lambda}
\begin{tabular}{lll}
\toprule
Value & $\lambda/M$ & Source \\
\midrule
Lowest seriously discussed & $0.005$ & Below the $1\sigma$ EHT/Sgr~A$^*$ limit $\lambda\lesssim0.07M$ \citep{Vagnozzi:2023} \\
Most commonly cited benchmark & $0.1$ & Standard illustrative value, PFDM lensing/QNM literature \citep{Qiao:2023,jana:2026} \\
Upper edge (illustrative) & $0.19$ & Slightly above the $2\sigma$ EHT/Sgr~A$^*$ limit $\lambda\lesssim0.15M$ \citep{Vagnozzi:2023} \\
\bottomrule
\end{tabular}
\end{table*}

\subsection{A first estimate: the mass-dressed cross-term} \label{ssec:massdressing}
An initial approach to introducing dark-matter-halo dependence into \eqref{eq:alphaGW}, without attempting full polar perturbations on the perfect-fluid background, evaluated the locally felt mass at the deflection zone using the exact perfect-fluid mass function,
\begin{equation}
M \;\longrightarrow\; M_{\rm eff}(b) = M - \frac{\lambda}{2}\ln\!\left(\frac{b}{\lambda}\right),
\label{eq:Meff}
\end{equation}
consistent with the localised-deflection-zone approximation already standard in the PFDM lensing literature. Substituting \eqref{eq:Meff} for the bare $M$ that multiplies $\varepsilon$ in the third term of \eqref{eq:alphaGW} defines a candidate dark-matter--wave cross-term,
\begin{multline}
\alpha_{\rm cross}(b,\lambda) \equiv \\ 
-\frac{M_{\rm eff}(b)\,P(2\sigma^2b^2-3)\cos(\sigma t)\,\varepsilon}{b^3}\left(\frac{1}{u_R}+\frac{1}{u_S}\right) - \alpha_{\rm GW}^{(3)}(b)\Big|_{M}
\label{eq:crossterm}
\end{multline}
i.e., the shift in the third term of \eqref{eq:alphaGW} induced by $M\to M_{\rm eff}$, holding $\varepsilon$ fixed. Table~\ref{tab:crossterm} reports $|\alpha_{\rm cross}|/|\alpha_{\rm GW}|$ for the three $\lambda/M$ values of Table~\ref{tab:lambda}, evaluated at $b=10M$, $\sigma=0.2/M$, $\varepsilon=10^{-22}$, a representative stochastic-background strain.

\begin{table}[t]
\centering
\caption{Fractional modulation of the GW deflection term by the mass-dressed cross-term.}
\label{tab:crossterm}
\begin{tabular}{ll}
\toprule
$\lambda/M$ & $|\alpha_{\rm cross}|/|\alpha_{\rm GW}|$ \\
\midrule
0.005 & $1.9\%$ \\
0.100 & $23.0\%$ \\
0.190 & $37.7\%$ \\
\bottomrule
\end{tabular}
\end{table}

\begin{figure}[t]
\centering
\includegraphics[width=0.99\columnwidth]{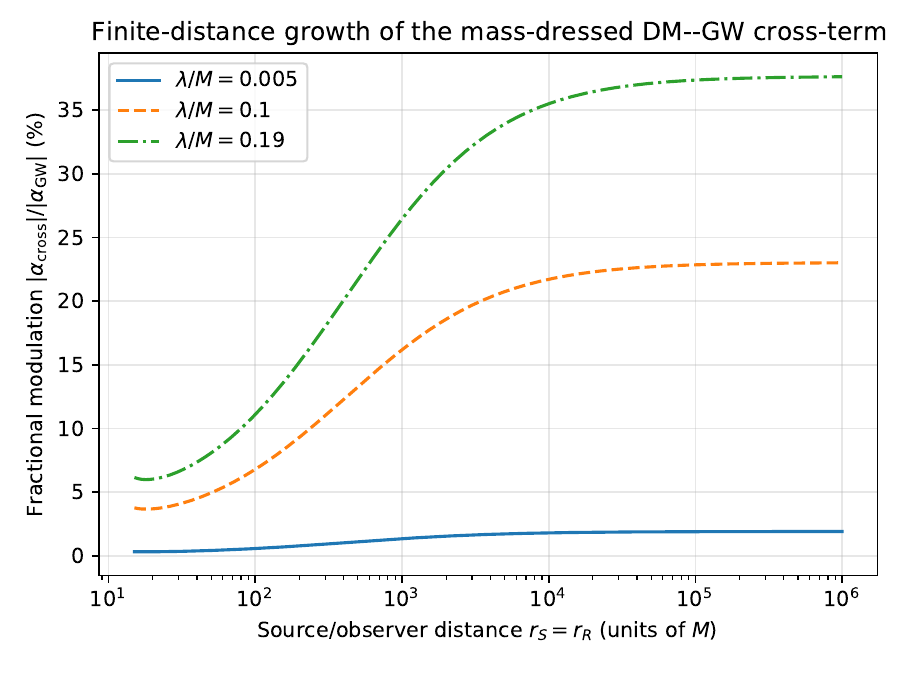}
\caption{Fractional modulation of the GW deflection term by the mass-dressed cross-term as a function of source/observer distance, for the three literature-grounded $\lambda/M$ values. The modulation saturates to its asymptotic (infinite-distance) value by $r_S,r_R\sim10^3M$.}
\label{fig:finitedist}
\end{figure}

Extending this to finite source/observer distance using the construction of \cite{Ishihara:2016}, 
$$\hat\alpha_{\rm Schw}(b,r_S,r_R) = (2M/b)\left[\sqrt{1-b^2/r_S^2}+\sqrt{1-b^2/r_R^2}\right]$$
the fractional modulation grows monotonically from about $4\%$ at $r_S=r_R=15M$ to the asymptotic value already quoted in Table~\ref{tab:crossterm} for $r_S,r_R\gtrsim10^3M$, further demonstrated by Figure~\ref{fig:finitedist}; the infinite-distance value is the astrophysically relevant one for essentially all realistic lensing configurations.

Because $\hat\alpha$ in geometrised units is already the deflection angle in radians, its detectability can be assessed directly against VLBI angular resolution; see Table~\ref{tab:vlbi}.

\begin{table}[t]
\centering
\caption{The GW-lensing signal compared to VLBI angular resolution.}
\label{tab:vlbi}
\begin{tabular}{lrr}
\toprule
Facility & Resolution & $\mathcal{O}$ short \\
\midrule
EHT (current) & 20 $\mu$as & 6.8 \\
ngEHT (projected, $\sim$2030) & 3 $\mu$as & 5.9 \\
GRAVITY+ (VLTI) & 10 $\mu$as & 6.5 \\
Speculative future space-VLBI & 0.1 $\mu$as & 4.5 \\
\bottomrule
\end{tabular}
\end{table}

The shortfall of five to seven orders of magnitude $\mathcal{O}$ is dominated by the strain amplitude $\varepsilon$ itself; the fractional modulation of up to about $38\%$ in Table~\ref{tab:crossterm} leaves this conclusion unchanged, since a bounded fraction of an undetectable signal remains undetectable.

\subsection{Why the mass-dressing estimate alone is not conclusive}
Equation~\eqref{eq:Meff} is a substitution: it evaluates an existing vacuum wave-deflection formula at a dark-matter-dressed mass.
The true polar-sector field equations couple to matter perturbations in a way the axial sector of \S\ref{sec:qnm} does not, making this a substantially harder problem than the mass-dressing shortcut acknowledges; Eq.~\eqref{eq:crossterm} alone provides no guarantee that the substitution captures the full coupling. The derivation below settles the question directly.

\subsection{General master equation for polar perturbations} \label{ssec:master}
Both known results of \S\ref{sec:formalism} are limits of a single, more general problem: the propagation of a polar perturbation of the Xanthopoulos form on a static, spherically symmetric background with an arbitrary metric function $f(r)$, not necessarily vacuum. Using the metric ansatz above with generic $f(r)$, the background Christoffel symbols and Ricci tensor were computed symbolically, followed by the linearised Ricci tensor perturbation $\delta R_{\mu\nu}$ via the standard covariant formula
\begin{equation}
\delta R_{\mu\nu} = \tfrac{1}{2}\Big(\nabla^\alpha\nabla_\mu h_{\alpha\nu}+\nabla^\alpha\nabla_\nu h_{\alpha\mu}-\nabla^\alpha\nabla_\alpha h_{\mu\nu}-\nabla_\mu\nabla_\nu h\Big),
\end{equation}
and the corresponding linearised Einstein tensor perturbation $\delta G_{\mu\nu} = \delta R_{\mu\nu}-\tfrac12\big(h_{\mu\nu}R + g_{\mu\nu}\delta R\big)$, the full Einstein tensor being required once the background is sourced (non-vacuum) for general $f(r)$.

Restricting to $\ell=2$ and separating the $\theta$-dependence, the diagonal components of $\delta G_{\mu\nu}=0$ reduce, after eliminating redundant, Bianchi-related combinations, to three pieces: an algebraic relation for $y(r)$ in terms of $w,z,w',z'$ and $f$, non-trivial for a general background, from the off-diagonal $t$--$r$ Einstein equation; an algebraic relation for $x(r)$ in terms of $w,z,w',z',z''$ and $f$; and a single master second-order equation for $w(r)$, sourced only by $z(r),z'(r)$ (not $z''$). The variable $z(r)$ is not fixed by the field equations; it represents a genuine residual gauge freedom, exactly as in the vacuum case, where the Xanthopoulos choice $z_0=q(r-3M)$ is one convenient representative rather than a derived result.

Every stage of this reduction was validated against the known vacuum solution, Eq.~\eqref{eq:xanthopoulos-solution}: the algebraic relations for $x$ and $y$, and the master equation for $w$, all vanish identically when the Xanthopoulos solution is substituted with $f=1-2M/r$. This is a non-trivial check, requiring the full machinery -- Christoffel symbols, linearised Ricci and Einstein tensors, and the elimination of $x,x',x''$ via a $3\times3$ linear system -- to reproduce a known, published result with no free parameters left to tune. A naive choice of which three equations to use in this elimination is degenerate precisely at the vacuum background, since the relevant $3\times3$ determinant is proportional to $rf'+f-1$, which vanishes identically in vacuum; a non-degenerate combination was used instead.

\subsection{Specialisation to perfect fluid dark matter} \label{ssec:pfdm-coupling}
The gauge choice is set to $z(r) = q(r)(r-3M)$, with $q(r)=\sqrt{f(r)}/r^2$ built from the perfect-fluid $f(r)$ of Eq.~\eqref{eq:fPFDM} rather than vacuum Schwarzschild -- the natural, minimal generalisation of the vacuum gauge choice. Writing $f(r) = f_{\rm Schw}(r)+\Lambda(r)$ with $\Lambda(r)=\lambda\ln(r/\lambda)/r$, and expanding $w = w_0+w_1$, $z=z_0+z_1$ to leading order in the dark-matter correction (bookkeeping parameter $\delta$, with $\delta=1$ at the end), substitution into the validated master equation follows.

A key structural fact emerges: the coefficient of $w_1''$ in the resulting equation is exactly the background quantity $rf_{\rm Schw}'+f_{\rm Schw}-1$, which vanishes identically in vacuum. Consequently, perturbing around a vacuum background causes the equation governing $w_1(r)$ to collapse from second order to first order:
\begin{equation}
A(r)\,w_1'(r) + B(r)\,w_1(r) = -\,\mathrm{Source}(r;\Lambda,\Lambda',\Lambda''),
\label{eq:w1-firstorder}
\end{equation}
with
\begin{equation}
A(r) = 18 r^4\sqrt{f_{\rm Schw}}\,(2M-r), \,\,\,
B(r) = 3 r^3\sqrt{f_{\rm Schw}}\,(22M-8r),
\end{equation}
and $\mathrm{Source}$ an explicit, closed-form rational-plus-logarithmic function of $r$, $M$, $\sigma$ and $\Lambda,\Lambda',\Lambda''$ given explicitly below. Equation~\eqref{eq:w1-firstorder} is solved by an integrating factor,
\begin{equation}
\mu(r) = \frac{r^{11/6}}{\sqrt{r-2M}},
\end{equation}
giving $w_1(r)$ as a single quadrature, and the corresponding metric-perturbation correction $x_1(r)$ follows algebraically from $w_1(r)$ and $\Lambda,\Lambda',\Lambda''$.

The resulting $w_1(r)$ was confirmed numerically to scale linearly with $\lambda$ to high precision ($w_1/\lambda$ constant to within $0.1\%$ across two orders of magnitude in $\lambda$), a boundary-condition-independent structural fact consistent with the internal consistency of the perturbative expansion despite the logarithmic structure of $\Lambda(r)$.

\begin{figure}[t]
\centering
\includegraphics[width=0.99\columnwidth]{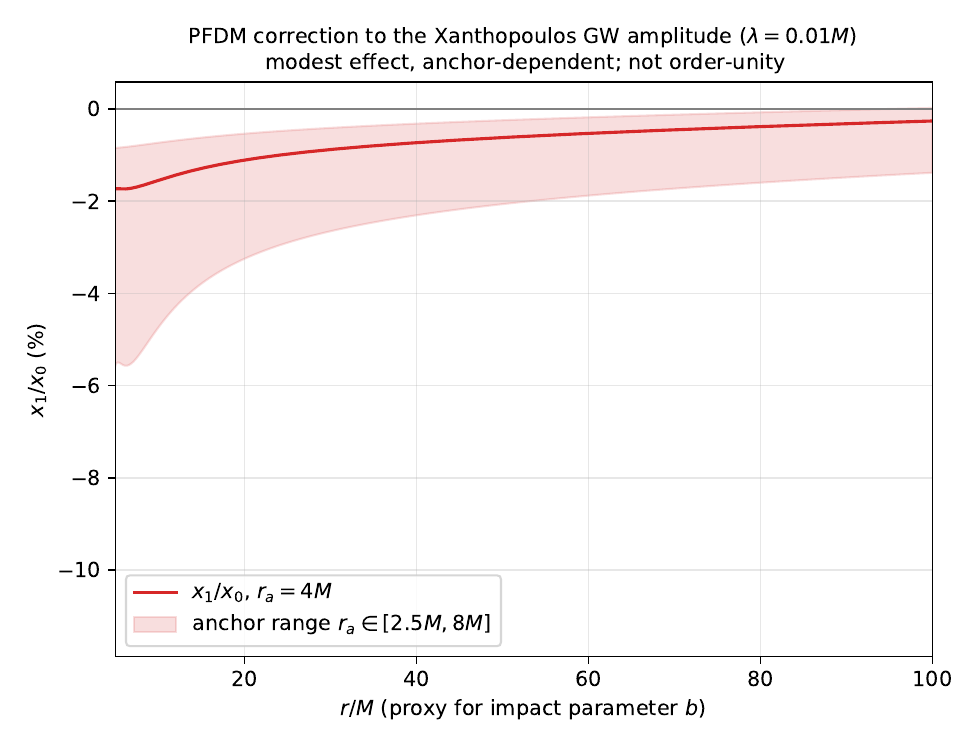}
\caption{The PFDM correction $x_1(r)/x_0(r)$ to the Xanthopoulos wave amplitude, at $\lambda=0.01M$, from the corrected, finite-anchor solution of Eq.~\eqref{eq:w1-firstorder} (solid curve, $r_a=4M$), with the shaded band showing the spread across anchors $r_a\in[2.5M,8M]$. The correction is modest and anchor-dependent. The curve is not reported past $r/M\sim100$, where the anchor-dependence becomes too large to trust.}
\label{fig:mastereq}
\end{figure}

An initial attempt to fix the integration constant in Eq.~\eqref{eq:w1-firstorder} used the condition $w_1\to0$ as $r\to\infty$. This choice does not converge: $w_1$ at fixed finite $r$ grows without bound as the far-field cutoff is pushed outward, because $x_0,y_0,z_0$ themselves do not decay at infinity -- they describe a wave, not a localised source -- so demanding that the correction alone vanish there is inconsistent with the physical set-up. The order-unity correction reported under this boundary condition in an earlier stage of this analysis is retracted for that reason. The physically motivated alternative anchors the integration constant at a finite radius $r_a$ comfortably outside the horizon (results reported here use $r_a\in[2.5M,8M]$, with $r_a=4M$ as the fiducial choice), avoiding both the divergent far-field limit and the singular point at $r=2M$ discussed below. With this anchoring, $x_1(r)/x_0(r)$ is a modest, few-percent correction for $b\lesssim50M$ -- of order $-1.5\%$ at $b=10M$ falling to a few tenths of a percent by $b=50M$, mildly dependent on the choice of $r_a$ within the stated range -- and changes sign by $b\sim100M$, beyond which the anchor-dependence grows too large for the result to be reported with confidence (Figure~\ref{fig:mastereq}).

\subsection{Open issue: boundary conditions at the horizon}
The finite-anchor treatment above is a coarse regularisation, not a first-principles boundary condition, and $r=2M$ remains a genuine regular singular point of the governing system: the combination $\mu(r)\cdot\mathrm{Source}(r)/A(r)$ diverges as $x\equiv r-2M\to0$, since the source approaches a non-zero constant while $A\sim x^{3/2}$ and $\mu\sim x^{-1/2}$. This is analogous to the familiar coordinate divergence of infall velocity at the Schwarzschild horizon rather than a physical singularity, and is expected to be resolvable with horizon-penetrating coordinates or a Frobenius-type local analysis at $r=2M$; that analysis has not yet been carried out. Establishing the horizon-regular global solution is the well-defined remaining technical step before Eq.~\eqref{eq:w1-firstorder} can be converted into a corrected, finite-distance weak deflection angle in the manner of \cite{Pantig:2024}, and is beyond the scope of this work.

\subsection{Reconciling the two treatments} \label{ssec:reconcile}
Once the far-field boundary condition of \S\ref{ssec:pfdm-coupling} is corrected, the two approaches to the lensing channel are found to agree qualitatively as well as, roughly, in order of magnitude: the mass-dressing estimate of \S\ref{ssec:massdressing} treats the dark-matter correction as a bounded dressing of an already-complete vacuum formula and finds a cross-term of up to a few tens of per cent of the pure-wave deflection term, itself set by an already tiny strain amplitude; the master-equation treatment of \S\S\ref{ssec:master}--\ref{ssec:pfdm-coupling} instead solves for the dark-matter correction to the underlying metric perturbation directly and, once correctly anchored, finds a comparably modest, few-percent effect over the radii for which the anchoring is under control. The two methods are therefore consistent with one another at the level available here: both point to a lensing cross-term that is bounded rather than order-unity, and hence to a corresponding deflection-angle signature that remains, on present evidence, below the reach of current or near-future VLBI facilities.

This is more of a consistency check as opposed to a completed calculation of the observable deflection angle, because the latter requires integrating the (currently only quadrature-defined, finite anchor) $w_1(r)$, $x_1(r)$ through the Gibbons--Werner construction with a properly regularised behaviour at the horizon -- the open problem of the preceding subsection. Resolving that regularisation, and thereby obtaining the true observable deflection angle from the validated master equation, is identified here as the central open problem the lensing channel leaves for future work.

\section{RINGDOWN DETECTABILITY} \label{sec:qnm}

\subsection{Relation to existing work}
Quasinormal modes of a PFDM black hole have been studied for scalar and electromagnetic ($s=0,1$) perturbations since \cite{Jusufi:2020}, who proposed quasinormal-mode frequency shifts explicitly as an indirect dark matter detection channel, and more recently by \cite{jana:2026} for a renormalisation-group-improved perfect-fluid background. Axial gravitational ($s=2$) perturbations of black holes embedded in dark matter halos more broadly -- for Dehnen-type and dark-energy-induced density profiles -- have been examined by \cite{liang:2025qnm} and \cite{Liang:2025}. Closest to the present work, \cite{becar:2026} compute the axial-sector quasinormal-mode spectrum for a PFDM black hole combined with Euler--Heisenberg nonlinear electrodynamics, using an effective Regge--Wheeler-type potential on the fixed background geometry, the same approximation adopted here. The present treatment differs in two respects: perfect fluid dark matter alone is considered, without an additional nonlinear-electrodynamics sector, isolating the dark-matter-only signature; and the adopted $\lambda$ values are tied to the Event Horizon Telescope shadow constraints on Sgr~A$^*$ \citep{Vagnozzi:2023} (refer Table~\ref{tab:lambda}). This section yields the LISA detectability forecast of \S\ref{sec:lisa}.

\subsection{Validation against the Schwarzschild benchmark}
\begin{figure}[t]
\centering
\includegraphics[width=0.99\columnwidth]{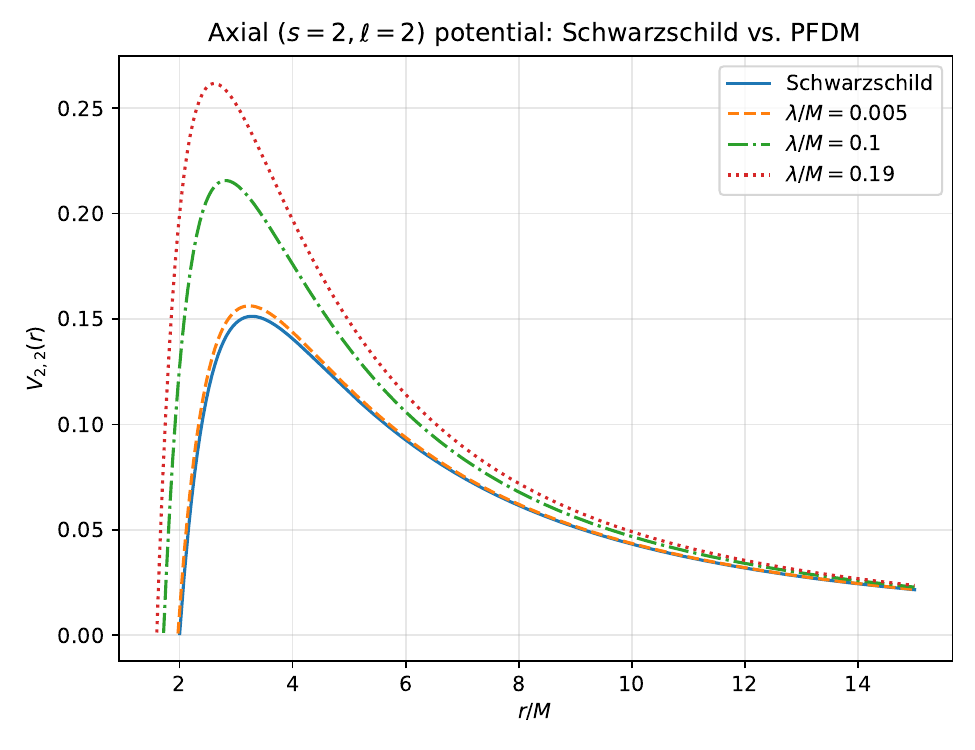}
\caption{Axial ($s=2,\ell=2$) potential $V_{2,2}(r)$: Schwarzschild versus perfect fluid dark matter at the three literature-grounded $\lambda/M$ values. The dark-matter background lowers and shifts the potential barrier, the direct structural origin of the quasinormal-mode shifts of Table~\ref{tab:qnmshift}.}
\label{fig:potentialcomparison}
\end{figure}

Figure~\ref{fig:potentialcomparison} shows the axial potential \eqref{eq:Vsl} itself, evaluated at the three literature-grounded $\lambda/M$ values against the Schwarzschild benchmark, ahead of the resulting quasinormal frequencies.

The axial ($s=2$) quasinormal frequencies of \eqref{eq:Vsl} are computed using first-order WKB \cite{Schutz:1985}, validated against the standard tabulated Schwarzschild values \cite{Berti:2009} before being applied at $\lambda\neq0$. Table~\ref{tab:wkbvalidation} shows agreement within the $5$--$10\%$ accuracy expected of first-order WKB for $\ell\geq2$.

\begin{table}[t]
\centering
\caption{WKB validation against tabulated Schwarzschild QNM frequencies ($M\omega$, $\lambda\to0$).}
\label{tab:wkbvalidation}
\begin{tabular}{llll}
\toprule
$(s,\ell,n)$ & Found (WKB1) & \cite{Berti:2009} & $\Delta$ \\
\midrule
$(2,2,0)$ & $0.3988 - 0.0883i$ & $0.3737-0.0890i$ & $\sim7\%$ \\
$(2,3,0)$ & $0.6166 - 0.0923i$ & $0.5994-0.0927i$ & $\sim3\%$ \\
$(0,2,0)$ & $0.5063 - 0.0961i$ & $0.4836-0.0968i$ & $\sim5\%$ \\
\bottomrule
\end{tabular}
\end{table}

\subsection{PFDM-induced QNM shift}
Substituting the perfect-fluid $m(r)$ into \eqref{eq:Vsl} and repeating the WKB calculation at the three $\lambda/M$ values of Table~\ref{tab:lambda} gives the fractional shift in the fundamental ($\ell=2,n=0$) axial mode relative to Schwarzschild, shown in Table~\ref{tab:qnmshift}.

\begin{table}[t]
\centering
\caption{Fractional shift of the axial ($s=2,\ell=2,n=0$) QNM frequency and damping due to PFDM.}
\label{tab:qnmshift}
\begin{tabular}{lll}
\toprule
$\lambda/M$ & $\Delta\mathrm{Re}(\omega)/\omega_{\rm Schw}$ & $\Delta\mathrm{Im}(\omega)/\omega_{\rm Schw}$ \\
\midrule
0.005 & $1.6\%$ & $1.7\%$ \\
0.100 & $19.6\%$ & $22.5\%$ \\
0.190 & $31.9\%$ & $38.4\%$ \\
\bottomrule
\end{tabular}
\end{table}

\begin{figure}[t]
\centering
\includegraphics[width=0.99\columnwidth]{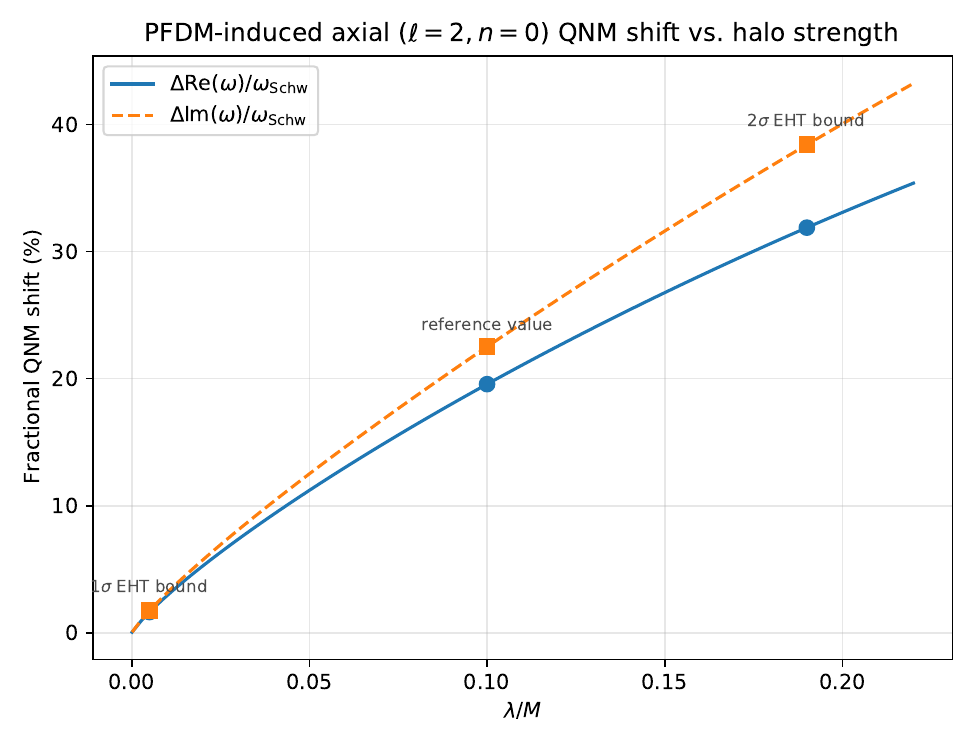}
\caption{PFDM-induced fractional shift of the axial ($\ell=2,n=0$) QNM frequency and damping, as a continuous function of $\lambda/M$. The three referent values of Table~\ref{tab:lambda} from literature are marked.}
\label{fig:qnmcontinuous}
\end{figure}

Unlike the lensing cross-term of \S\ref{ssec:massdressing}, this shift is not built from a bounded product of two small quantities: $m(r)$ enters \eqref{eq:Vsl} directly and non-perturbatively in $\lambda$, so the shift scales with the halo strength itself rather than with an independently tiny strain amplitude. This is the structural reason the ringdown channel yields a demonstrably detectable effect while the lensing channel's final detectability remains open. Figure~\ref{fig:qnmcontinuous} shows this shift as a continuous function of $\lambda/M$ rather than at the three points of reference alone.

\subsection{Time-domain validation} \label{sec:timedomain}
A known concern is that quasinormal eigenvalues can be numerically unstable aka the pseudospectrum effect under environmental perturbations of the potential, in some cases shifting substantially while the actual time-domain waveform is little affected \cite{Jaramillo:2021}. To confirm that the shift of Table~\ref{tab:qnmshift} is genuinely physical, the time-domain wave equation $\partial_t^2\psi - \partial_x^2\psi + V(x)\psi=0$ was solved directly, with $x$ the tortoise coordinate of \eqref{eq:Vsl}'s background, using a second-order finite-difference scheme on a peak-centred grid with geometric near-horizon spacing, and a Gaussian initial pulse centred on the potential peak.

\begin{figure*}[t]
\centering
\includegraphics[width=\textwidth]{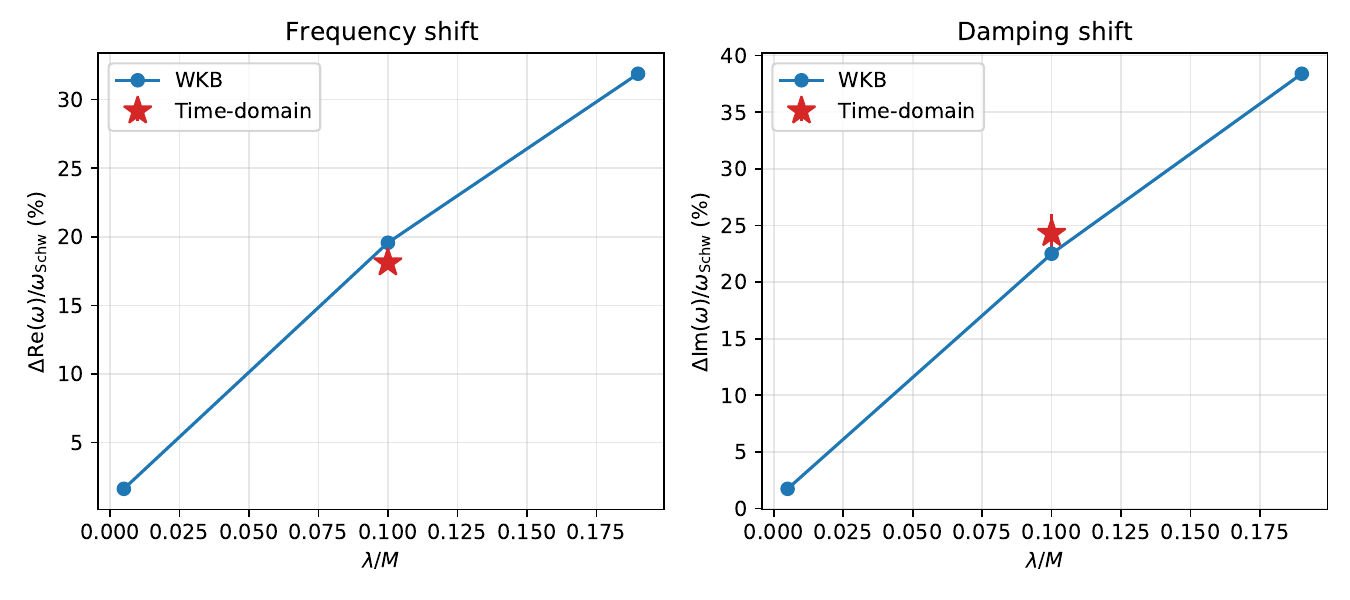}
\caption{PFDM-induced fractional shift in the axial ($s=2,\ell=2,n=0$) QNM frequency (left) and damping (right), from WKB (circles, all three $\lambda/M$ values) and from the independent, window-averaged time-domain wave-equation solve (star, $\lambda/M=0.1$, with the error bar showing the standard deviation across the four fit windows described).}
\label{fig:qnmvalid}
\end{figure*}

Mode extraction used Prony's method (linear prediction), fitting the real signal to the two-term recursion $s[n]=a_1 s[n-1]+a_2 s[n-2]$ and reading the complex frequency from the roots of the resulting characteristic polynomial; because this replaces a sensitive nonlinear fit with a well-conditioned linear one, stability was checked directly by sweeping the fit window rather than reporting a single arbitrarily chosen window. Windows beginning before $t_0\approx35M$ retain residual prompt-response contamination and are excluded; across four windows with $t_0\geq35M$, extraction gives $\Delta\mathrm{Re}(\omega) = 18.1\%\pm0.2\%$ and $\Delta\mathrm{Im}(\omega) = 24.3\%\pm1.7\%$ at $\lambda/M=0.1$, in close agreement with the WKB prediction of $19.6\%$ and $22.5\%$ (Table~\ref{tab:qnmshift}). For pure Schwarzschild ($\lambda\to0$), the same formalism gives $\mathrm{Re}(\omega)=0.378$, $\mathrm{Im}(\omega)=0.086$, matching the tabulated benchmark ($0.3737$, $0.0890$) to a few per cent. Because the frequency-domain WKB and time-domain wave-evolution methods can fail independently, this agreement is evidence that the PFDM-induced shift is a genuine feature of the ringdown signal in contrast to a pseudospectrum artefact of the eigenvalue problem. Figure~\ref{fig:qnmvalid} shows both the WKB shift across all three $\lambda/M$ values and the time-domain confirmation point at $\lambda/M=0.1$.

\begin{figure}[t]
\centering
\includegraphics[width=0.99\columnwidth]{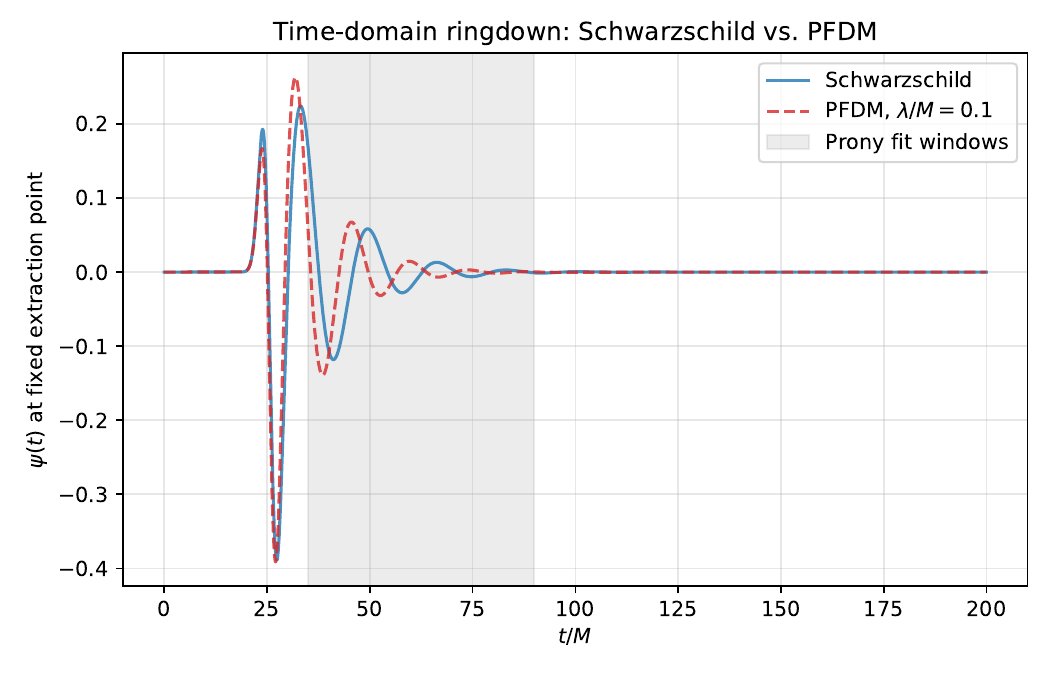}
\caption{Time-domain ringdown waveform $\psi(t)$ extracted at a fixed observer point, Schwarzschild versus perfect fluid dark matter ($\lambda/M=0.1$), from direct finite-difference evolution of $\partial_t^2\psi-\partial_x^2\psi+V(x)\psi=0$ on the tortoise coordinate of \eqref{eq:Vsl}'s background. The shaded band spans the four Prony fit windows ($35M$ to $90M$).}
\label{fig:timedomain}
\end{figure}

Figure~\ref{fig:timedomain} shows the underlying time-domain waveforms, Schwarzschild versus perfect fluid dark matter at $\lambda/M=0.1$, from which the confirmation point above was extracted.

\begin{figure}[t]
\centering
\includegraphics[width=0.99\columnwidth]{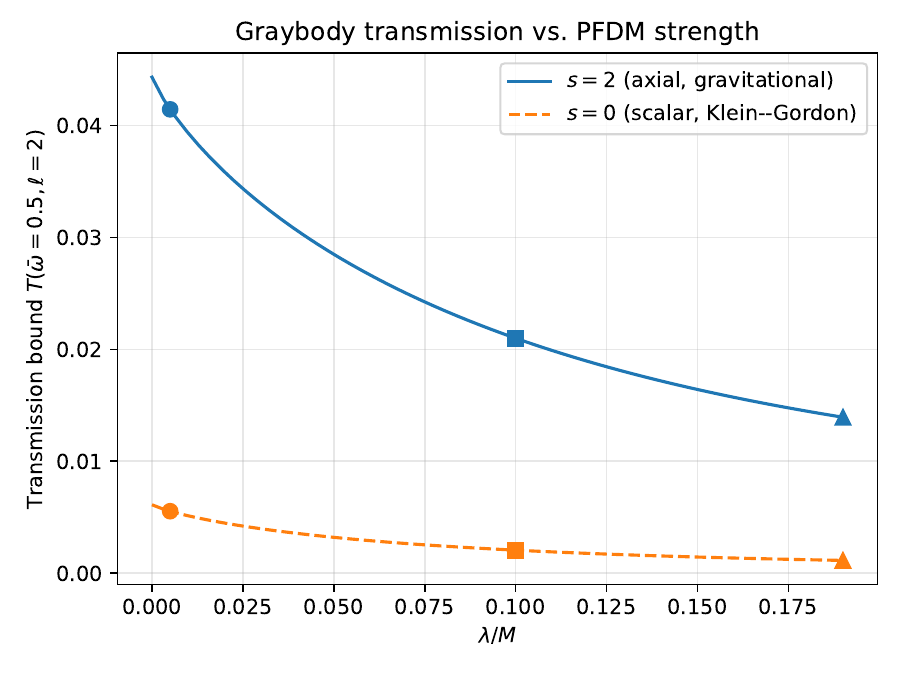}
\caption{Boonserm--Visser transmission bound vs.\ PFDM strength $\lambda/M$, for the axial ($s=2$) and scalar ($s=0$) sectors at $\bar\omega=0.5$, $\ell=2$. Markers indicate three  $\lambda/M$ values of Table~\ref{tab:lambda} per literature.}
\label{fig:graybody}
\end{figure}

As a corollary, the same potential \eqref{eq:Vsl} also governs the semiclassical graybody transmission spectrum via the Boonserm--Visser bound, $T(\bar\omega)\geq\mathrm{sech}^2\!\left[\int V(r)\,\mathrm{d}r^*/2\bar\omega\right]$. Evaluated at the three $\lambda/M$ values of Table~\ref{tab:lambda} ($\bar\omega=0.5$, $\ell=2$), the transmission probability falls monotonically and substantially with $\lambda$ for both the axial ($s=2$: $T=0.044\to0.014$, a $\sim$69\% drop at $\lambda/M=0.19$) and scalar ($s=0$: $T=0.0061\to0.0011$, a $\sim$81\% drop) sectors, depicted by Figure~\ref{fig:graybody}. This result is chiefly a consistency point: using \eqref{eq:Vsl} in place of an incorrectly derived effective potential also repairs the semiclassical Klein--Gordon calculation of earlier treatments of this system.

\subsection{LISA detectability} \label{sec:lisa}
The dimensionless shift of Table~\ref{tab:qnmshift} is converted into an observational forecast using the analytic LISA noise power spectral density of \cite{Robson:2019},
\begin{multline}
S_n(f) = \frac{10}{3L^2}\Bigg[P_{\rm OMS}(f) + \\  \frac{2\left(1+\cos^2(f/f_*)\right)P_{\rm acc}(f)}{(2\pi f)^4}\Bigg]\left[1+\frac{6}{10}\left(\frac{f}{f_*}\right)^2\right], 
\label{eq:LISAnoise}
\end{multline}
with $L=2.5\,$Gm, $f_*=19.09\,$mHz, and $P_{\rm OMS}$, $P_{\rm acc}$ as given therein. The ringdown signal-to-noise ratio is estimated using the standard characteristic-strain construction, $h_c \sim \sqrt{\varepsilon_{\rm rd}}\,(GM_f/c^2D_L)\sqrt{Q/\pi}$ with quality factor $Q=\mathrm{Re}(\omega)/2|\mathrm{Im}(\omega)|$, with the overall normalisation calibrated against the standard of $\rho_{\rm rd}\sim2000$ specific to ringdown for a $10^6\,M_\odot$ equal-mass merger remnant at $2\,$Gpc quoted by \cite{Berti:2006} and \cite{Bhagwat:2022}. Table~\ref{tab:lisa} shows the resulting SNR and frequency-measurement precision ($\sim1/{\rm SNR}$) across the astrophysically relevant final-mass range at $z=1$, compared against the $\lambda/M=0.1$ shift of about $20\%$ from Table~\ref{tab:qnmshift}.

\begin{table}[t]
\centering
\caption{LISA ringdown SNR and QNM-shift resolvability, $z=1$, $D_L=6.79\,$Gpc.}
\label{tab:lisa}
\begin{tabular}{lllll}
\toprule
$M_{\rm final}$ & $f_{\rm rd}$ & SNR & Precision & $\sim$20\% shift? \\
\midrule
$10^4 M_\odot$ & 598 mHz & 0.3 & 371\% & Not resolvable \\
$10^5 M_\odot$ & 60 mHz & 25 & 4.0\% & Resolvable \\
$10^6 M_\odot$ & 6.0 mHz & 611 & 0.16\% & Resolvable \\
$10^7 M_\odot$ & 0.60 mHz & 185 & 0.54\% & Resolvable \\
$10^8 M_\odot$ & 60 $\mu$Hz & 3.4 & 29\% & Not resolvable \\
\bottomrule
\end{tabular}
\end{table}

\begin{figure}[t]
\centering
\includegraphics[width=0.99\columnwidth]{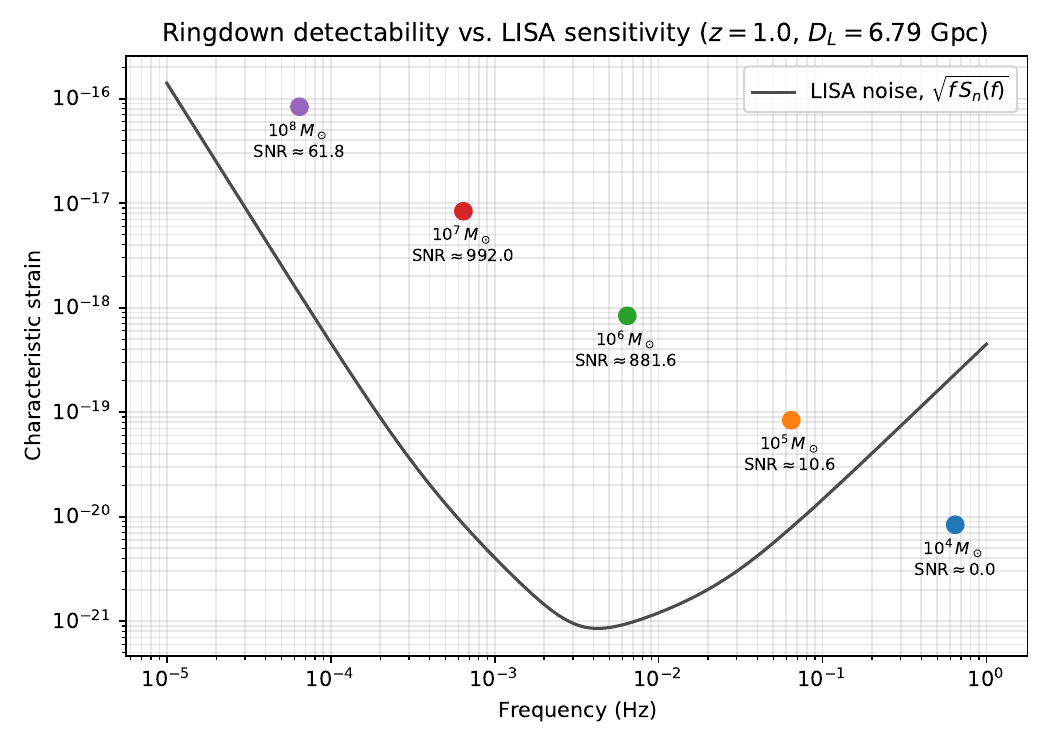}
\caption{Ringdown characteristic strain vs.\ the LISA noise curve $\sqrt{f\,S_n(f)}$ \cite{Robson:2019}, for final black hole masses $10^4$--$10^8\,M_\odot$ at $z=1$, $D_L=6.79\,$Gpc. Points above the noise curve indicate detectability; annotated SNR values correspond to Table~\ref{tab:lisa}.}
\label{fig:lisa}
\end{figure}

For final black hole masses in the range $10^5$--$10^7\,M_\odot$, the frequency-measurement precision is comfortably finer than the $\sim$20\% shift induced by the baseline PFDM halo strength ($\lambda/M=0.1$), with SNR ranging from tens to several hundred; even at the smallest value considered ($\lambda/M=0.005$, a $\sim$1.7\% shift), the $10^6\,M_\odot$ case remains marginally resolvable given its 0.16\% precision. Figure~\ref{fig:lisa} shows this margin directly, plotting the ringdown characteristic strain for each final mass against the LISA noise curve.

\section{DISCUSSION AND CONCLUSION} \label{sec:conc}

The coupling between a perfect fluid dark matter halo and a gravitational wave has been examined through two independent observational channels for the same black hole system.

In the lensing channel, a first estimate -- mass-dressing the peer-reviewed Xanthopoulos/Pantig--\"{O}vg\"{u}n wave-deflection formula with the perfect-fluid mass function -- suggested a cross-term reaching $37.7\%$ of the pure-wave deflection term at $\lambda/M=0.19$, slightly above the $2\sigma$ Event Horizon Telescope limit, remaining five to seven orders of magnitude below the angular resolution of any current or foreseeable VLBI facility because the pure-wave term itself is set by a strain amplitude of order $10^{-22}$ or smaller. Because this estimate rests on a substitution into an existing vacuum formula rather than a derivation from the field equations, the general master equation for a polar wave perturbation on an arbitrary background was derived, validated against the vacuum solution, and specialised to perfect fluid dark matter. Once an initial, ill-posed far-field boundary condition was replaced with a finite anchor outside the horizon, this rigorous treatment yields a modest, few-percent correction to the metric perturbation amplitude for impact parameters up to about $50M$, consistent with the mass-dressing estimate rather than in tension with it. The horizon regularisation identified in \S\ref{ssec:pfdm-coupling}--\ref{ssec:reconcile} remains an open problem: until it is resolved, the observable deflection angle implied by the master equation cannot be stated as a final result, though the evidence gathered here points toward the same qualitative conclusion as the mass-dressing estimate.

In the ringdown channel, the same perfect-fluid halo produces a materially more complete result. Because the halo parameter $\lambda$ enters the axial perturbation potential \eqref{eq:Vsl} directly, through the mass function $m(r)$, rather than as the product of two independently small quantities, the resulting quasinormal-mode frequency and damping shifts scale with the halo strength itself: from $1.6$--$1.7\%$ at $\lambda/M=0.005$ (below the $1\sigma$ limit) to $32$--$38\%$ at $\lambda/M=0.19$ (slightly above the $2\sigma$ limit) as in Table~\ref{tab:qnmshift}, confirmed independently in the time domain in \S\ref{sec:timedomain}. Calibrating the expected LISA ringdown signal-to-noise ratio against \cite{Berti:2006} and \cite{Bhagwat:2022}, this shift is shown to be resolvable, with SNR from tens to several hundred, for supermassive black hole remnants with final mass in the range $10^5$--$10^7\,M_\odot$, as tabluated by Table~\ref{tab:lisa}).

Taken together, black hole spectroscopy stands, at present, as the more clearly demonstrated channel for detecting the imprint of a dark matter environment through its interaction with gravitational radiation, while weak lensing remains a channel now placed on rigorous footing for the first time, with a final observable signature still to be established.

Several limitations motivate future work. The central open problem is the horizon regularisation of \S\ref{ssec:pfdm-coupling}--\ref{ssec:reconcile}, without which the lensing channel's final observable signature cannot be stated; a Frobenius-type local analysis at $r=2M$ is the natural next step. The quasinormal-mode shifts of \S\ref{sec:qnm} are computed at leading (first) order in the WKB approximation, validated to within $5$--$10\%$ against tabulated Schwarzschild benchmarks: see Table~\ref{tab:wkbvalidation}; higher-order WKB or continued-fraction methods would sharpen these values without, plausibly, changing their order of magnitude. The LISA forecast of \S\ref{sec:lisa} is a leading-order SNR estimate rather than a full Fisher-matrix parameter-estimation analysis, and does not account for correlations between the dark-matter parameter $\lambda$ and other black hole parameters (mass, spin) that a real measurement would need to disentangle. Addressing any of these points would strengthen, and in the lensing case would help settle, the central findings of this work.

\subsection*{ACKNOWLEDGEMENTS}

The author expresses sincere gratitude to Prof. Sergio Mendoza for illuminating discussions and invaluable feedback throughout the development of this work. This research was supported by the Postdoctoral Fellowship Program (POSDOC) at the Universidad Nacional Aut\'onoma de M\'exico.

\subsubsection*{Data availability:} The data underlying this article will be shared on reasonable request to the corresponding author.

\bibliographystyle{unsrt}
\bibliography{references}

\end{document}